\documentclass[lettersize,journal]{IEEEtran}
\usepackage{amsmath,amsfonts}
\usepackage{algorithmic}
\usepackage{algorithm}
\usepackage{array}
\usepackage[caption=false,font=normalsize,labelfont=sf,textfont=sf]{subfig}
\usepackage{textcomp}
\usepackage{stfloats}
\usepackage{url}
\usepackage{verbatim}
\usepackage{graphicx}
\usepackage{cite}
\usepackage{changepage}

\usepackage{tabularx}
\usepackage[most]{tcolorbox}
\usepackage{threeparttable}
\usepackage{booktabs}
\usepackage{nicematrix}
\PassOptionsToPackage{hyphens}{url}
\usepackage{hyperref}
\usepackage{makecell}
\usepackage{booktabs}
\usepackage{siunitx}
 \usepackage{fontawesome}
\usepackage{pifont}
\usepackage{enumerate}
\usepackage{enumitem}
\usepackage{graphicx}
\usepackage{changepage}

\usepackage{pifont}
\usepackage{caption}
\usepackage{subcaption}
\usepackage{minted}

\usepackage{colortbl}
\usepackage[table,xcdraw]{xcolor}

\newcommand{\tool}{\textsf{VerifiaBLE}\xspace}

\usepackage{xspace}
\usepackage{listings}
\usepackage{multirow}

\usepackage{booktabs}
\definecolor{bluetoothBlue}{RGB}{0, 0, 255}
\definecolor{hexRed}{RGB}{180, 0, 0}
\definecolor{commentGray}{RGB}{128, 128, 128}

\usepackage[table]{xcolor}
\usepackage{colortbl}
\usepackage{xintexpr, xinttools}

\newcommand{\ccell}[3]{%
    \cellcolor{pink!\xinttheiexpr 90*#1/#2\relax}#3%
}

\begin{document}

\title{\textit{What You ``Code'' Is What We ``Prove''}:\\ Translating BLE App Logic into Formal Models with LLMs for Vulnerability Detection}

\author{Biwei~Yan,~\IEEEmembership{}
        Yue~Zhang,~\IEEEmembership{}
        Minghui~Xu,~\IEEEmembership{}
        Runyu~Pan,~\IEEEmembership{}
        Jinku~Li,~\IEEEmembership{Member,~IEEE},
        Xiuzhen~Cheng,~\IEEEmembership{Fellow,~IEEE}

\thanks{
B. Yan, Y. Zhang, M. Xu, R. Pan, and X. Cheng are with the School of Computer Science and Technology at Shandong University, China. Jinku Li is with the School of Cyber Engineering at Xidian University, China. (e-mail: \{bwyan, zyueinfosec, mhxu, rypan, xzcheng\}@sdu.edu.cn, jkli@xidian.edu.cn). (\textit{Corresponding author: Yue Zhang.})
}

\thanks{
This work was partially supported by the National Natural Science Foundation of China (Grant No.62402288, 62402291), the China Postdoctoral Science Foundation 2024M751811, National Key Research and Development Program of China (Grant No.2022YFB4502001).
}
}

\markboth{Journal of \LaTeX\ Class Files,~Vol.~14, No.~8, August~2021}%
{Shell \MakeLowercase{\textit{et al.}}: A Sample Article Using IEEEtran.cls for IEEE Journals}


\maketitle

\begin{abstract}
The application layer of Bluetooth Low Energy (BLE) is a growing source of security vulnerabilities, as developers often neglect to implement critical protections such as encryption, authentication, and freshness. While formal verification offers a principled way to check these properties, the manual effort of constructing formal models makes it impractical for large-scale analysis. This paper introduces a key insight: BLE application security analysis can be reframed as a semantic translation problem, i.e., from real-world code to formal models. We leverage large language models (LLMs) not to directly detect vulnerabilities, but to serve as translators that convert BLE-specific code into process models verifiable by tools like ProVerif. We implement this idea in VerifiaBLE, a system that combines static analysis, prompt-guided LLM translation, and symbolic verification to check three core security features: encryption, randomness, and authentication. Applied to 1,050 Android BLE apps, VerifiaBLE uncovers systemic weaknesses: only 10.2\% of apps implement all three protections, while 53.9\% omit them entirely.  Our work demonstrates that using LLMs as structured translators can lower the barrier to formal methods, unlocking scalable verification across security-critical domains.
\end{abstract}


\section{Introduction}
\IEEEPARstart{B}{luetooth} Low Energy (BLE) has become the de facto communication protocol for a wide range of smart devices, including wearables, medical monitors, smart lighting, and home automation systems~\cite{zuo2019automatic,zhang2020bless}. While the BLE standard specifies a well-defined protocol stack, from radio communication through the attribute and security layers, the application layer remains largely unconstrained. Developers are responsible for defining custom services, characteristics, and data flows on top of the protocol. This design flexibility fosters innovation but also introduces significant and often overlooked security risks. Although BLE supports secure pairing at lower layers, many apps still transmit sensitive data or accept control commands without applying encryption, authentication, or freshness checks at the application layer. This results in real-world risks such as eavesdropping, spoofing, and replay attacks. \looseness=-1

However, most BLE security research targets protocol~\cite{wu2020blesa,zhang2020breaking} or firmware flaws~\cite{wen2020firmxray}, yet the application layer (where developers define custom data flows) is both underexamined and often the source of the most critical vulnerabilities. Analyzing security at this layer is particularly challenging due to the absence of standardized specifications~\cite{woolley2019bluetooth} and the highly heterogeneous nature of real-world implementations. For instance, one app may transmit authentication tokens via custom characteristics, while another may use unstructured binary payloads with implicit control semantics. Prior efforts such as \textit{BLEScope}~\cite{zuo2019automatic} and \textit{BLESS}~\cite{zhang2020bless} have made initial progress by tracking cryptographic API usage or control-flow patterns, but their coverage is limited to shallow syntactic features and predefined rules. They often miss nuanced security logic (e.g., whether a nonce is actually linked to a sensitive operation), and as a result, current tools struggle to capture the full semantics of BLE interactions.
In principle, formal verification provides a powerful and principled solution to this challenge.   However, the use of formal methods in practice is severely limited by their steep usability cost: crafting accurate formal models requires expert knowledge, manual effort, and precise abstraction of real-world behaviors. This barrier makes formal verification largely inaccessible for large-scale analysis of BLE applications.

At the core of this challenge is a translation problem: BLE applications implement communication logic in imperative code, while formal tools reason over abstract process models (or known as modeling language). Bridging this semantic gap, i.e., converting real-world BLE byte code into formal modeling language, is the critical missing link. Recognizing this, we reframed BLE application verification as a code-to-code translation task: given code that defines how data is sent, encrypted, or authenticated over BLE, can we automatically generate a corresponding formal modeling language suitable for verification?
This perspective naturally points to large language models (LLMs)~\cite{vaswani2017attention,li2024attention,yao2024survey} as an enabling component. While LLMs are not reliable for directly detecting vulnerabilities, they excel at understanding programming semantics and generating well-structured output in other representations \cite{10606356}. We therefore explore a design where LLMs serve not as vulnerability analyzers, but as semantic translators that convert BLE communication logic into ProVerif-compatible models. This allows us to leverage the flexibility of LLMs for extraction and generation, while delegating soundness and judgment to symbolic verification.

To this end, we present \tool (formal \textbf{Verifi}cation \textbf{a}nalysis for \textbf{BLE} applications), a system that integrates BLE-specific static analysis with retrieval-augmented prompt engineering to extract relevant code paths, translate them into formal process models via LLMs, and verify them using \textsf{ProVerif}. 
We evaluate VerifiaBLE on 1,050 real-world BLE Android applications and conduct extensive manual validation and case studies. Our findings reveal systemic weaknesses in BLE application-layer security:  our tool revealed that only 10.2\% implemented all required core security features (e.g., encryption, randomness, authentication), while 53.9\% without any protections. We observed that secure design adoption strongly correlates with app popularity and quality: high-download and high-rated apps are more likely to implement BLE protections, though gaps remain. For example, 23.5\% of apps with over 1M downloads still lacked freshness guarantees, making them vulnerable to replay attacks. Category-level analysis revealed that apps in Health, Business, and Medical domains were more likely to adopt encryption and authentication, whereas Education, Game, and Social apps remained largely insecure.  Developer-level analysis showed consistent security behavior across apps by the same publisher, with some consistently secure (e.g., Facebook) and others consistently insecure. Over 60\% of apps with multiple versions showed no improvement in BLE security over time, suggesting long-term neglect of BLE protocol hygiene in app maintenance.

This paper makes the following contributions:

\begin{itemize}
    \item We propose a novel architecture that uses LLMs as semantic translators to bridge unstructured BLE logic and formal verification tools. Translating real-world code into formal models offers a practical bridge between software engineering and formal verification; this insight may generalize to other domains. 
    \item We design and implement \tool, combining static BLE-specific analysis, prompt-guided model generation, and symbolic verification of three core security properties.\looseness=-1
    \item We conduct the largest empirical study to date on BLE application-layer security, revealing widespread flaws and validating them through formal analysis and real-device attacks.
\end{itemize}

\section{Background}\label{secII}

\subsection{Bluetooth Low Energy and Its Security}

\begin{figure}[!tb]
	\centering 
	\includegraphics[width=0.37\textwidth]{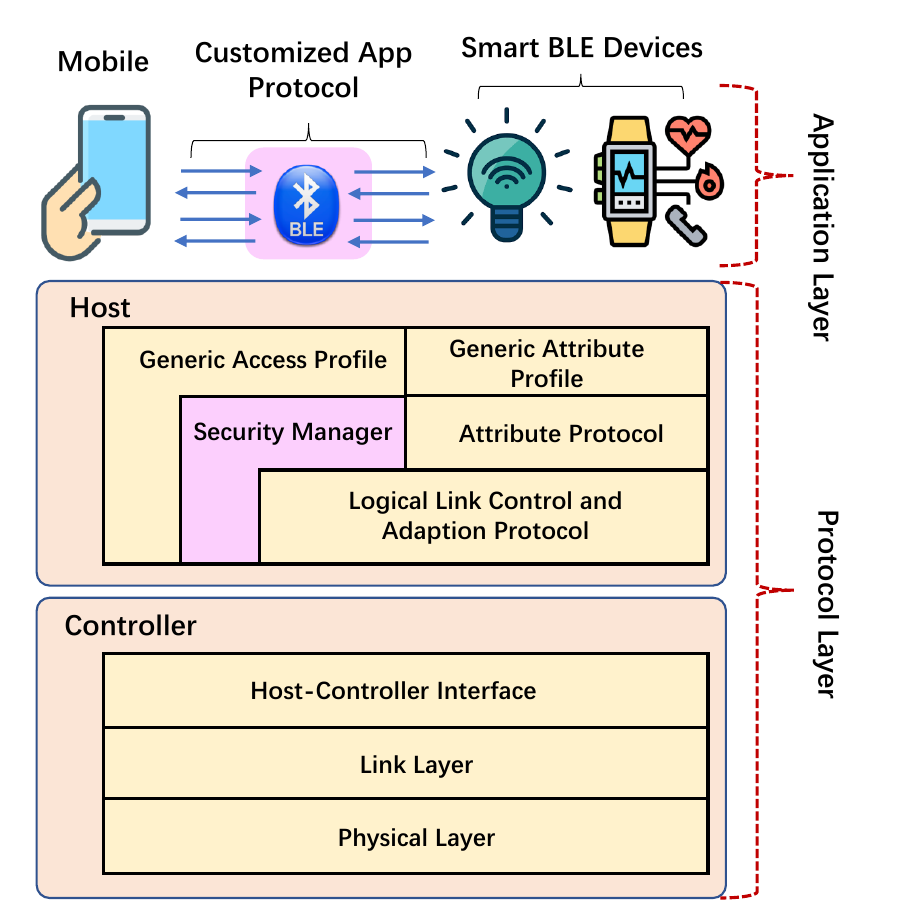}
	\caption{BLE Protocol Stack and Security Architecture}
	\label{fig:blestack}
\end{figure}

\vspace{1mm}
\noindent\textbf{BLE Communication and Stack.}
BLE is a widely adopted low-power wireless communication protocol used in IoT devices such as smart locks, fitness trackers, and beacons. In a typical BLE connection, a central device (e.g., smartphone) scans for nearby peripherals that periodically broadcast advertisement packets containing device and service information. Once a match is found, the central initiates a connection, followed by optional pairing involving key exchange and authentication to enable secure communication. 
\autoref{fig:blestack} illustrates the BLE protocol architecture, which includes a protocol layer and an application layer. The protocol layer comprises the host (GAP, GATT, ATT, L2CAP, Security Manager) and controller (physical/link layers, HCI). While the protocol layer handles standardized communication and security features, the application layer remains flexible, where developers define services and data flows atop the stack.

The application layer includes the customized application protocol (e.g., smart light, smart watch) running between the mobile app and the BLE device. This part is typically implemented by developers to meet specific application needs. While the BLE stack already includes built-in security mechanisms, developers may implement additional security features such as application-layer encryption, mutual authentication, or integrity checks within this customized protocol to further protect sensitive data and ensure secure communication tailored to their use case. The red-bracketed region in the figure highlights where core BLE security is implemented, within both the application layer the protocol layer.  

\vspace{1mm}
\noindent\textbf{BLE Security.} At the protocol layer, BLE pairing assumes physical proximity during the bonding phase, treating physical access as an implicit trust anchor. For example, in Passkey Entry mode, a BLE smart lock displays a temporary code (e.g.,  ``\textit{123456}'') that must be manually entered into the phone—ensuring that only someone physically near the lock can pair with it. While this mechanism enables link-layer encryption, it does not prevent impersonation: an attacker who observes the code (e.g., in public spaces) can pair a rogue device and gain persistent control. Because the protocol lacks device-level identity checks post-pairing, BLE systems that rely solely on link-layer protection remain vulnerable.  For example, a publicly accessible BLE smartlock using passkey entry authentication can be compromised when an attacker physically views and enters the displayed passcode into their smartphone. Once paired, the lock's reliance solely on link-layer security grants the attacker complete control, as the system cannot distinguish between authorized and malicious paired devices. To address these vulnerabilities,   BLE-based applications may implement additional security at the application layer (e.g., challenge-response mechanisms, certificate-based validation).  By enforcing strong authentication before granting access, BLE devices can prevent unauthorized control, even if an attacker successfully establishes a link-layer connection. 

\subsection{Formal Verification and Its BLE Applications}

Formal verification rigorously proves whether a system satisfies security properties like confidentiality, integrity, and authentication across all possible executions. ProVerif~\cite{930138}, based on applied pi-calculus, is a widely used tool for verifying cryptographic protocols through symbolic analysis. Its scalability to unbounded sessions makes it suitable for complex IoT protocols such as BLE. Given the security-critical and resource-constrained nature of IoT devices, formal methods like ProVerif have been effectively used to analyze BLE pairing, authentication, and privacy, uncovering flaws beyond the reach of conventional testing.

To use Formal verification such as ProVerif for analyzing protocols (e.g., a BLE protocol), testers begin by abstracting the protocol’s communication flows into a formal model using the applied pi-calculus. This involves defining the roles of participants (e.g., initiator and responder), the cryptographic primitives used (e.g., symmetric encryption, public key operations, nonces), and the message exchanges. The security properties to be verified (such as secrecy of session keys or mutual authentication) are then specified as queries. For example, a confidentiality query may check whether a secret key can be derived by an attacker. Once the model is complete, ProVerif automatically analyzes all possible protocol executions to determine whether the specified properties hold. If a property fails, ProVerif provides a trace showing how the violation could occur, enabling developers to identify and patch the flaw.



\section{ Problem Statement and Key Idea}\label{sec3}

\subsection{Motivation and Problem Statement}

\vspace{2mm}
\noindent\textbf{Motivation.}
Although traditional BLE vulnerability detection methods have made significant strides, particularly in protocol-layer detection through string matching or traffic analysis such as identifying \texttt{device.createBond} in Android's implementations, they still face fundamental limitations. A key challenge lies in detecting security flaws at the application layer, where developers often implement custom protocols without standardized formats. This diversity makes it difficult for automated tools to generalize detection logic. \looseness=-1

In response to these challenges, some researchers have turned to program analysis techniques in an effort to evaluate application-layer security more effectively. However, these approaches often rely on assumptions that may not hold in practice. For example, \textit{BLEScope}~\cite{zuo2019automatic} assume that secure BLE apps typically derive cryptographic keys from user input rather than hard-coded values in the code. Their taint analysis tracks whether user-provided data flows into BLE communication requests; if so, the app is deemed secure. However, this assumption is not always reliable. User input may not actually be used to construct keys, and even if it is, the resulting implementation may still be insecure due to flawed logic or a misunderstanding of security principles. For instance, as shown in \autoref{fig:ble-code}, in the case of a real-world smart light app called \textsf{AppLights} (which we manually analyzed and tested), the core protection mechanism involved using the device's MAC address to ``\textit{encrypt}'' the communication payload. This design is insecure: since the MAC address is publicly available and transmitted in plaintext during each communication, an attacker can easily reverse the operation and recover the original message. 

\begin{figure}[h!]
\begin{adjustwidth}{2em}{0em} 
 
\begin{minted}[fontsize=\footnotesize, linenos]{java}
ArrayList localArrayList = new ArrayList<>();
while ((paramAnonymousArrayOfByte == null) || 
       (paramAnonymousArrayOfByte.length <= 8))
       {
    String[] arrayOfString = 
    paramAnonymousBluetoothDevice.
    getAddress().split(":");
    while (arrayOfString.length != 6) {
        int i = 0; if (i < 6) {break label969;
        }}}
while ((byte)(1 - (short)((short)
((localObject[0] & 0xFF) * 
        (localObject[2] & 0xFF)) 
        + (short)((localObject[3] & 0xFF) * 
        (localObject[5] & 0xFF))))) {
    Object localObject = new HashMap<>();
    ((Map)localObject).put("addr", str);
    ((Map)localObject).put("keyarray",
    localArrayList);}

\end{minted}
\end{adjustwidth}
\caption{Code snippet of \textsf{AppLights}}
\label{fig:ble-code}
\end{figure}

To evaluate the security of such custom BLE protocols more rigorously, formal verification techniques are often employed. Analysts typically begin by reviewing the application’s source code and abstracting the protocol logic into a formal specification language, such as applied pi-calculus. They then define security goals like confidentiality or authentication, and use tools such as ProVerif to verify whether these properties hold. While formal methods can uncover subtle flaws that are difficult to detect through empirical testing, the modeling process is manual and time-intensive, requiring both protocol expertise and familiarity with formal verification tools, factors that limit their widespread adoption in practice.

These limitations highlight the need for more practical, automated, and intelligent tools that can bridge the gap between low-level protocol analysis and high-level application logic. An ideal tool would combine the precision of formal verification with the scalability of static or dynamic analysis, while minimizing reliance on brittle assumptions or manual modeling.

\begin{table}[t]
\centering
\caption{Impact of missing core security features on categorized BLE attacks}
\scriptsize
\begin{tabular}{llccc}
\toprule
\setlength\tabcolsep{2pt}
\textbf{Attack} & \textbf{Category} & \textbf{w/o Enc.} & \textbf{w/o Nonce} & \textbf{w/o Auth.} \\
\midrule
Eavesdropping           & Passive         & $\checkmark$ & $\times$     & $\times$ \\
Traffic analysis        & Passive         & $\checkmark$ & $\times$     & $\times$ \\
Replay attack           & Active          & $\times$     & $\checkmark$ & $\times$ \\
Message injection       & Active          & $\times$     & $\checkmark$ & $\checkmark$ \\
Message modification    & Active          & $\times$     & $\checkmark$ & $\checkmark$ \\
Spoofing                & Active          & $\times$     & $\times$     & $\checkmark$ \\
Man-in-the-middle       & Active          & $\checkmark$ & $\checkmark$ & $\checkmark$ \\
\bottomrule
\end{tabular}
\label{tab:ble-systematic-attacks}
\end{table}

\vspace{2mm}
\noindent\textbf{Problem Statement and Scope.}
In this work, we aim to develop a static analysis tool that can reason about the security posture of BLE applications by identifying whether core communication security mechanisms are properly implemented at the application layer. While high-level objectives such as confidentiality and integrity remain foundational to security design, they are often too abstract to support automated or large-scale analysis. Instead, we focus on three concrete and measurable security features, i.e., encryption, randomness (e.g., nonces), and authentication, which operationalize these abstract goals in a form that is directly observable in application code.
This feature-centric methodology is not only technically grounded, but also reflects common practice~\cite{zhang2020bless,zuo2019automatic} in BLE and mobile application security research. As shown in \autoref{tab:ble-systematic-attacks}, there are:

\begin{itemize}
\item \textbf{Encryption} is necessary to protect sensitive communication content from passive adversaries. Applications that omit encryption leave transmitted data exposed to eavesdropping and behavioral inference. 

\item \textbf{Randomness (Nonces)} protects against replay and session-reuse attacks by ensuring message freshness. BLE applications that fail to use nonces can be vulnerable to manipulation via recorded message replays.  
\item \textbf{Authentication} mechanisms prevent impersonation and man-in-the-middle (MITM) attacks by verifying the identity of connected devices. Without authentication, an adversary can hijack or mimic BLE devices.  
\end{itemize}

We further scope our analysis to BLE applications on the Android platform. Android represents one of the most widely deployed platforms for BLE-based consumer applications, including smart home controllers, fitness trackers, medical monitors, and more. Additionally, we focus our analysis on applications whose code is not obfuscated. While obfuscation is a common practice for protecting intellectual property or complicating reverse engineering, it also introduces significant challenges for static or semantic analysis by hindering code structure recovery and control/data flow tracking.

 

\subsection{Key Idea}
We propose a hybrid analysis pipeline that combines static program analysis with LLM-assisted formalization. Our approach first extracts BLE-relevant communication logic from Android applications using static analysis, identifying key functions, control flows, and message construction paths. Then, we use LLMs to translate these code fragments into formal models, which are subsequently verified by tools like \textit{ProVerif}. 

Our key insight is to instead leverage LLMs as an assistant, not as a judge, to bridge the gap between real-world code and formal verification.
This architecture offers several advantages. First, the task of converting BLE communication related code into formal specifications can be naturally framed as a structured translation problem (mapping imperative code to a symbolic, declarative model). This framing aligns directly with the core strength of transformer-based large language models, which were originally developed for translation tasks.  
Second, it avoids relying on LLMs for direct vulnerability classification, mitigating issues of hallucination and non-determinism. 
While LLMs excel at tasks like vulnerability pattern recognition or code summarization, their predictions are often ungrounded and difficult to trust in isolation, especially when used as black-box classifiers. Directly asking an LLM ``\textit{is this code secure?}'' often yields inconsistent or overly optimistic responses due to the lack of symbolic reasoning and the difficulty of interpreting long-range dependencies in program semantics. Prior works~\cite{yin2024multitask, steenhoek2024err, ding2024vulnerability} have shown that directly using LLMs to classify or detect vulnerabilities often yields unreliable results, frequently no better than random guessing.


\begin{figure*}[!tb]
	\centering 
	\includegraphics[width=1\textwidth]{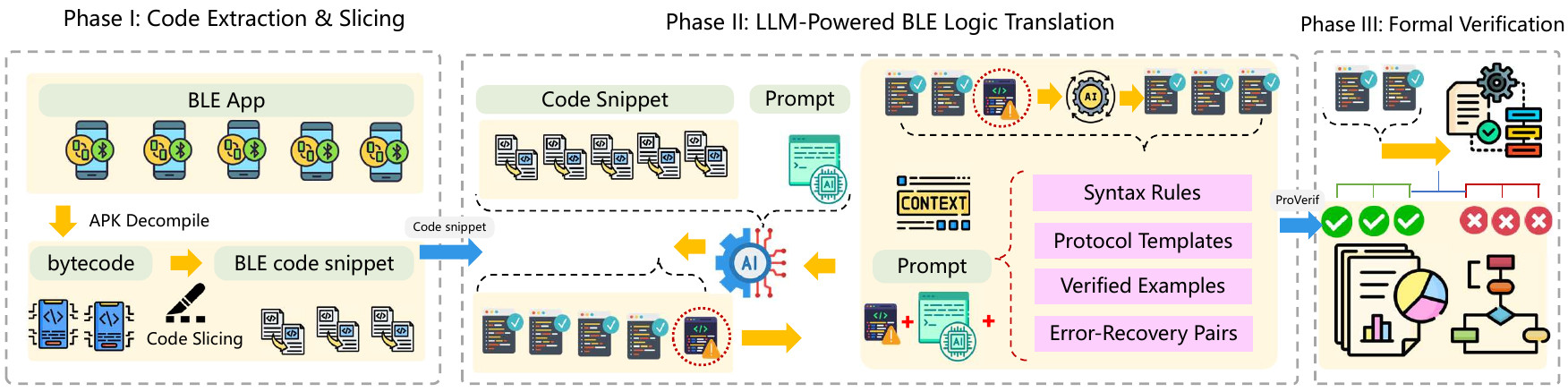}
	\caption{Design of \tool:  {Phase I extracts BLE code snippets from APKs. Phase II uses an LLM and RAG context to generate ProVerif models. Phase III verifies these models and reports potential vulnerabilities.}}
	\label{detail}
\end{figure*}

\section{Design of \tool} \label{sec4}

In this section, we present \tool,  formal \textbf{Verifi}cation \textbf{a}nalysis for \textbf{BLE} applications, powered by LLM-guided semantic modeling. The overall architecture of \tool is illustrated in ~\autoref{detail}, and the entire pipeline is organized into three distinct phases:

\begin{itemize}
    \item   \textbf{Code Extraction and Slicing.} We first decompile the APK to recover Java byte code.  Given the token limitations of LLMs, processing entire codebases directly is impractical. To address this, we adopt a code slicing strategy that isolates smaller, semantically meaningful snippets relevant to BLE operations. Specifically, we perform static analysis on the decompiled code to identify functions interacting with BLE APIs. \looseness=-1
     \item   \textbf{LLM-Powered BLE Logic Translation.}  In this phase, we translate the filtered BLE-related code into formal specifications using a large language model enhanced with a retrieval-augmented generation (RAG) framework. To ensure correctness, we employ chain-of-thought prompting and few-shot learning to guide the LLM in producing well-structured \textsf{ProVerif} code. Security queries (e.g., those verifying confidentiality or authentication) are embedded in the output.\looseness=-1

        \item \textbf{Formal Verification.}
In the final phase, we conduct formal verification using the \textsf{ProVerif} models generated in Phase II. These models encode the BLE communication logic and associated security queries derived from the application code.   If the verification engine identifies violations, it produces symbolic attack traces that illustrate how an adversary could exploit flaws in the application’s BLE logic. \tool parses this trace and generates a vulnerability report that explicitly identifies the type of attack.  
\end{itemize}

{The output of each phase flows directly into the next. The BLE-related code snippets generated in the first phase form the core input for the ProVerif code generation in the second phase. Combined with the context retrieved by RAG, the LLM is guided to produce correct and complete ProVerif code. These ProVerif models are then used by the verifier in the third phase to output vulnerability results.}

\subsection{Code Extraction and Slicing}






To analyze BLE security at the code level, we adopt a two-step process: first, we reconstruct Java byte code from APKs to recover high-level logic; second, we extract BLE-relevant code by identifying API usage and recursively tracing dependencies. 

\vspace{1mm}
\noindent\textbf{(Step-I) Code Extraction.}
This step aims to transform binary Android applications into readable Java byte code, thereby enabling subsequent code analysis, slicing, and formal verification.  We begin by parsing the APK file to extract its core contents, focusing on retrieving the Dalvik Executable (DEX) files that contain the compiled bytecode. These DEX files are then converted into Java Archive (JAR) format to enable compatibility with standard Java decompilers. The resulting JAR files are processed to generate human-readable Java byte code, capturing class structures, methods, and inter-method dependencies.

\vspace{1mm}
\noindent\textbf{(Step-II) Code Slicing.}
Once the Java code is reconstructed, the next step is to isolate the program logic relevant to BLE communication. This is necessary because BLE-related security mechanisms such as connection setup, authentication, and data transmission, are often interwoven with general-purpose application logic.  Extracting only the relevant code not only improves the accuracy of formal verification. 
If irrelevant or incomplete code is passed to the LLM, the generated formal models may be incorrect or incomplete, leading to false positives or missed vulnerabilities in the verification phase.  {Since LLMs process input tokens at a fixed computational cost, including unrelated code introduces semantic noise and increases token overhead. This may degrade performance, reduce output interpretability, and even cause model truncation or timeout in complex cases.}

To accurately isolate BLE communication logic from decompiled Android applications, we first analyze the reconstructed class and method hierarchy. The primary goal is to identify program regions that directly or indirectly involve BLE functionality, such as device discovery, service binding, and characteristic read/write operations. This step is particularly important in Android apps, where BLE interactions are often fragmented across multiple lifecycle methods, callbacks, and utility functions, making manual identification error-prone and incomplete.

We begin by scanning for known BLE API usages, including but not limited to method calls referencing \texttt{BluetoothAdapter}, \texttt{BluetoothDevice}, \texttt{BluetoothGatt}, and \texttt{BluetoothGattCallback}. These serve as entry points into the BLE interaction flow. Once the initial BLE-related methods are identified, we perform a recursive traversal of their call graph to retrieve all transitive dependencies. This ensures that helper methods (such as those handling data formatting or asynchronous callbacks) are not excluded, even if they are several layers removed from the BLE API calls. For example, in a smart lock control app, the BLE write operation was abstracted through multiple utility methods, including a custom encryption wrapper. A naive keyword match would have overlooked these layers. By recursively traversing the call graph, we captured the full interaction chain, enabling a semantically complete and security-relevant code slice.


\subsection{LLM-Powered BLE Logic Translation}

This phase focuses on translating BLE-related application code into a formal representation executable by the \textsf{ProVerif} verifier. However, this translation task is inherently non-trivial: the source code is written in imperative Java with complex control structures and domain-specific abstractions, while \textsf{ProVerif} expects a declarative, process-oriented representation. Bridging this semantic gap requires careful prompt design, semantic abstraction, and iterative refinement. \looseness=-1

 

\vspace{1mm}
\noindent\textbf{(Step-I) Prompt-Guided Formalization via LLMs.}
We begin by transforming the extracted BLE-relevant code snippets into corresponding \textsf{ProVerif} models using LLMs. To improve generation accuracy, we design specialized prompts incorporating both chain-of-thought (CoT) reasoning and few-shot learning. The prompts instruct the LLM to reason step-by-step about the security-relevant logic such as message exchanges, key derivation, and identity checks—and then map it to formal constructs like \texttt{process}, \texttt{event}, and \texttt{query}. For instance, when analyzing a BLE pairing function that performs mutual authentication via passkey exchange, the LLM can infer a corresponding model involving fresh nonce generation, private channel establishment, and a secrecy query for the passkey.

\vspace{1mm}
\noindent\textbf{(Step-II) Self Error Detection via RAG.} Despite careful prompt design, the generated \textsf{ProVerif} code may contain syntactic or semantic errors that prevent successful verification. These issues arise due to subtle grammar mismatches, incomplete translation of logic branches, or misuse of process composition operators, which challenges that are amplified by the formal language's strict parser. 
To automatically detect  errors in the initially generated formal models, we construct a domain-specific \textsf{ProVerif} knowledge base and integrate it into a RAG pipeline. This knowledge base serves as both a semantic memory and a structured reference, enabling the LLM to retrieve grounded, context-aware examples and rules that guide the correction process.  The knowledge base comprises four key components:
\begin{itemize}
\item \textbf{Syntax Rules}: Specifies the grammar of \textsf{ProVerif}, including constructs like \texttt{let}, \texttt{if-then-else}, \texttt{in}, \texttt{new}, \texttt{event}, and \texttt{query}, helping the LLM avoid structural errors such as incorrect block placement or malformed declarations.

\item \textbf{Protocol Templates}: Provides verified models of common cryptographic protocols (e.g., challenge-response, ephemeral key exchange), abstracting reusable patterns such as nonce exchange and mutual authentication used in BLE pairing and bonding.

\item \textbf{Best Practices and Examples}: A curated set of high-quality \textsf{ProVerif} code from academic sources and case studies, demonstrating idiomatic usage and domain-specific strategies like modeling out-of-band key exchange or session resumption.

\item \textbf{Error-Recovery Pairs}: A repository of real-world code generation failures and their corrections (e.g., misplaced conditionals or misordered events), enabling the LLM to learn from past mistakes and generate repairable code.
\end{itemize}

\vspace{1mm}
\noindent\textbf{(Step-III) Iterative Repair and Verification.}
The retrieved context is then incorporated into a new prompt, along with the original faulty code and its error trace, to guide the LLM in generating a corrected version. For example, if the model mistakenly defines a public channel for a key exchange process, the retrieved fix may include a correctly scoped \texttt{private channel} example. The generation process explicitly references syntactic constructs and semantic expectations from the knowledge base, reducing hallucinations and ensuring that the output adheres to \textsf{ProVerif}'s specification. The newly generated code is re-submitted to \textsf{ProVerif} for verification. If errors persist or security properties are not satisfied, the  ``retrieve-generate-verify''  loop continues until either a valid formal model is synthesized or a maximum retry threshold is reached. This automated loop reduces the need for human-in-the-loop correction.\looseness=-1

\subsection{Formal Verification}

Once a syntactically correct and semantically meaningful \textsf{ProVerif} model is generated, we submit it to the \textsf{ProVerif} tool to perform formal verification of the application's BLE communication logic. This step serves as the final checkpoint in our analysis pipeline to determine whether the modeled protocol satisfies essential security requirements under a symbolic attacker model. In our work, we focus on verifying the presence and correctness of three key security features, i.e., encryption, randomness (nonces), and authentication.

\textsf{ProVerif} supports the formal verification of a broad range of security properties. Among these, two classes of properties are most commonly used in practice: secrecy and correspondence assertions. Secrecy queries (e.g., \texttt{query attacker(x)}) are used to express confidentiality properties.  
To model and verify the use of randomness, particularly the presence of nonces that prevent replay attacks, we declare fresh names (e.g., \texttt{new n: bitstring;}) within the protocol model. We then analyze whether these nonces are reused or exposed in a way that could allow replay or correlation by an adversary. Correctly used nonces should remain unpredictable and session-specific, and their security is reflected in the absence of derivable attack traces involving reused messages. 
Authentication, including message integrity and origin verification, is expressed via correspondence assertions. These are specified using events such as \texttt{event begin\_auth(x)} and \texttt{event end\_auth(x)}, and encoded as logical implications like \texttt{query event(end\_auth(x)) ==> event(begin\_auth(x))}. Such assertions ensure that a critical protocol step (e.g., accepting a session key) only occurs if a legitimate initiating action took place. Violations of these assertions indicate potential vulnerabilities such as spoofing, man-in-the-middle attacks, or forged command execution. \looseness=-1

\section{Evaluation}\label{sec5}

\subsection{Experiment Setup}

To validate the feasibility of our approach, we conducted tests on over $1050$ BLE apps, which were collected from the AndroZoo database \cite{allix2016androzoo}. AndroZoo is a widely used repository of Android apps designed for researchers and developers, aimed at facilitating security research and malware analysis on the Android platform. The repository currently contains $8,054,736$ different APKs covering various application categories and provides an API interface to support bulk downloading. In this paper, we download $8296$ android apps, from which we filter $1050$ BLE apps. Therefore, we utilize BLE-related functions (such as \texttt{connectGatt} and \texttt{onLeScan}) as well as relevant permission settings (e.g., \texttt{android.permission.BLUETOOTH}) to filter out BLE apps. This approach allows researchers to locate and analyze functions related to BLE communication. 
In our experiments, we use GPT-4o as the core LLM to generate \textsf{ProVerif} models for vulnerability detection. 

\subsection{Performance of VerifiaBLE} \label{performance}
 
 
\noindent To evaluate the effectiveness of our \tool in detecting vulnerabilities in BLE applications, we conducted a manual analysis of the detection results. Specifically, we randomly sampled 100 applications flagged as secure and 100 applications flagged as insecure. For each sample, we manually inspected the BLE-relevant execution paths extracted by our tool, focusing on whether the app correctly implements three core security features: encryption, randomness (e.g., nonces), and authentication. 
Among the 100 apps marked as secure, 96 were confirmed to correctly implement all three features, while 4 were identified as false positives (FPs). Importantly, the main sources of FP stem from incomplete or ineffective implementations of security features that superficially appear correct. For example, some apps invoke encryption functions within the BLE logic, but use insecure cipher modes, rendering the protection ineffective. On the other hand, among the 100 apps flagged as insecure, all were correctly identified. The low error rates and explainable causes demonstrate that our tool provides a reliable and interpretable framework for large-scale BLE vulnerability detection.

 

\noindent To further evaluate \tool's practical effectiveness, we compare its detection accuracy against two state-of-the-art BLE security analysis frameworks: BLESS and BLECryptracer. BLESS is a static analysis tool that traces the use of cryptographic keys and nonces to identify potential BLE security flaws, while BLECryptracer assesses an app’s security posture based on the presence of cryptographic API calls in BLE-related code. 
Due to the lack of a widely adopted benchmark dataset with ground-truth security labels for BLE applications, we reused the same dataset discussed in RQ1—100 applications flagged as secure and 100 flagged as insecure. For each tool, we manually validated its classification results to determine accuracy.
As shown in ~\autoref{table_comparison}, BLESS and BLECryptracer correctly identified 75 and 7 secure applications, corresponding to precision rates of 86\% and 10\%, respectively. For insecure applications, BLESS achieved a perfect detection rate by correctly identifying all 100 apps, while BLECryptracer identified 94.
\tool outperforms both tools by achieving more balanced and higher accuracy in detecting both secure and insecure BLE applications.

\begin{table}[h]
    \centering
    \setlength\tabcolsep{3pt}
    \caption{Detection Accuracy Comparison Between Tools}
    \label{table_comparison}
    
    \begin{tabular}{@{}lcc@{}}
        \toprule
        \textbf{Tool} & \textbf{Secure Apps Acc.} & \textbf{Insecure Apps Acc.} \\
        \midrule
        BLESS~\cite{zhang2020bless} & 86\% & 100\% \\
        BLECryptracer~\cite{sivakumaran2019study} & 10\% & 94\% \\
        BLEScope~\cite{zuo2019automatic} & 88\% & 96\% \\
     \rowcolor{red!12}
       \tool & \textbf{96\%} & \textbf{100\%} \\
        \bottomrule
    \end{tabular}
 
\end{table}

\subsection{Experiment Results}
 {In Section \ref{performance}, we can ensure that \tool demonstrates strong performance in detecting BLE vulnerabilities. 
Subsequently, we use \tool to perform a large-scale static analysis of 1,050 Android BLE applications, aiming to uncover potential risks in real-world deployments.
}

\begin{center}
     \noindent\fbox{%
    \parbox{0.95\columnwidth}{%
 
      \textbf{(RQ1)} \textbf{What is the security distribution of BLE apps, including the number classified as secure and insecure?} 
    }%
}
\end{center}
\vspace{2mm}

\noindent 

To understand the current state of security practices in real-world BLE applications, we conducted a large-scale static analysis of 1,050 Android BLE apps. We systematically analyzed whether the application implemented any of the following three core security features:   encryption,   randomness and  authentication. 
As shown in~\autoref{table:security_combination}, we classified each application into one of eight categories.   
 Notably, the most common configuration, i.e., lacking all three features, appears in 566 applications (53.9\%), indicating that over half of the apps fail to implement even minimal security safeguards, leaving them open to passive and active attacks.
The second most frequent configuration, found in 308 apps (29.33\%), includes encryption and randomness but omits authentication. While such apps offer some protection against eavesdropping and replay attacks, the absence of peer verification makes them susceptible to impersonation and MitM attacks. 
Only 107 applications (10.19\%) implement all three features. This low proportion highlights a widespread neglect of authentication mechanisms, despite their importance for verifying device identity and preventing unauthorized access.



\begin{table}[htbp]
\centering
\scriptsize
\setlength\tabcolsep{6pt}
\caption{Combinations of BLE Security Features: Encryption, Nonce Usage, and Authentication}
\label{table:security_combination}
\begin{tabular}{cccrr}
\toprule
\textbf{Encryption} & \textbf{Nonce} & \textbf{Authentication} & \textbf{Num. (\#)} & \textbf{Percent (\%)} \\
\midrule
\checkmark & \checkmark & \checkmark & 107 & 10.19 \\
\checkmark & \checkmark & $\times$    & 308 & 29.33 \\
\checkmark & $\times$    & \checkmark & 5   & 0.48 \\
\checkmark & $\times$    & $\times$    & 56  & 5.33 \\
$\times$    & \checkmark & \checkmark & 1   & 0.10 \\
$\times$    & \checkmark & $\times$    & 3   & 0.29 \\
$\times$    & $\times$    & \checkmark & 4   & 0.38 \\
$\times$    & $\times$    & $\times$    & 566 & 53.90 \\
\bottomrule
\end{tabular}
\end{table}

\vspace{2mm}
\begin{center}
     \noindent\fbox{%
    \parbox{0.95\columnwidth}{%
 
      \textbf{(RQ2)} \textbf{How does the adoption of core BLE security mechanisms vary across different categories?} 
    }%
}
\end{center}
\vspace{2mm}
To explore how BLE security features are adopted across different application domains, we categorized the analyzed applications according to their functional types (e.g., Business, Education, Entertainment) and examined the presence of three core security mechanisms. \autoref{feature_combinations_by_category} presents the distribution of BLE application categories across all eight possible combinations of these features. 
We observe a clear disparity in the adoption of security mechanisms among different industry sectors, suggesting that certain application scenarios may systematically overlook specific types of protections. For example, \textit{Education} applications rarely incorporate authentication mechanisms; among all combinations, the variant with full security coverage (\textbf{E}ncryption, \textbf{N}once, and \textbf{A}uthentication) appears in less than 2\% of cases. Similarly, in \textit{Entertainment} and \textit{Game} applications, the use of the nonce mechanism, which is essential for preventing replay attacks,is extremely limited. This indicates that applications targeted at general consumers may prioritize functionality over robust security measures. 
In contrast, categories that handle sensitive user data, such as \textit{Business} and \textit{Health \& Lifestyle}, demonstrate a more balanced use of multiple security mechanisms. In particular, combinations involving both cryptography and authentication  occur more frequently. These applications tend to place greater emphasis on ensuring privacy and transaction security.

\begin{table}[htbp]
\centering
\scriptsize
\setlength{\tabcolsep}{1.4pt}
\caption{Distribution of BLE Application Feature Combinations by Category}
\begin{tabular}{lcccccccccccccccccccccccc}
\toprule
\multirow{2}{*}{\textbf{Categories}} & 
\textbf{E} & \textbf{N} & \textbf{A} & 
  \textbf{E} & \textbf{N} & \textbf{A} &
  \textbf{E} & \textbf{N} & \textbf{A} &
  \textbf{E} & \textbf{N} & \textbf{A} & 
  \textbf{E} & \textbf{N} & \textbf{A} & 
  \textbf{E} & \textbf{N} & \textbf{A} & 
  \textbf{E} & \textbf{N} & \textbf{A} & 
  \textbf{E} & \textbf{N} & \textbf{A}  \\
\cmidrule(lr){2-4} \cmidrule(lr){5-7} \cmidrule(lr){8-10} \cmidrule(lr){11-13} \cmidrule(lr){14-16} \cmidrule(lr){17-19} \cmidrule(lr){20-22} \cmidrule(lr){23-25}
& 
  \ding{55} & \ding{55} & \ding{55} & 
  \ding{55} & \ding{55} & \ding{51} &
  \ding{55} & \ding{51} & \ding{55} &
  \ding{55} & \ding{51} & \ding{51} & 
  \ding{51} & \ding{55} & \ding{55} & 
  \ding{51} & \ding{55} & \ding{51} & 
  \ding{51} & \ding{51} & \ding{55} & 
  \ding{51} & \ding{51} & \ding{51}\\
\midrule
Business     & \multicolumn{3}{c}{51.63\%} & \multicolumn{3}{c}{0.54\%} & \multicolumn{3}{c}{--} & \multicolumn{3}{c}{--} & \multicolumn{3}{c}{5.43\%} & \multicolumn{3}{c}{1.09\%} & \multicolumn{3}{c}{27.17\%} & \multicolumn{3}{c}{14.13\%} \\
Education  & \multicolumn{3}{c}{44.62\%} & \multicolumn{3}{c}{--} & \multicolumn{3}{c}{--} & \multicolumn{3}{c}{--} & \multicolumn{3}{c}{4.62\%} & \multicolumn{3}{c}{--} & \multicolumn{3}{c}{49.23\%} & \multicolumn{3}{c}{1.54\%} \\
Entertainment & \multicolumn{3}{c}{41.27\%} & \multicolumn{3}{c}{--} & \multicolumn{3}{c}{--} & \multicolumn{3}{c}{--} & \multicolumn{3}{c}{7.94\%} & \multicolumn{3}{c}{--} & \multicolumn{3}{c}{44.44\%} & \multicolumn{3}{c}{6.35\%} \\
Shopping        & \multicolumn{3}{c}{65.87\%} & \multicolumn{3}{c}{--} & \multicolumn{3}{c}{--} & \multicolumn{3}{c}{--} & \multicolumn{3}{c}{3.17\%} & \multicolumn{3}{c}{--} & \multicolumn{3}{c}{25.40\%} & \multicolumn{3}{c}{5.56\%} \\
Game                    & \multicolumn{3}{c}{51.19\%} & \multicolumn{3}{c}{--} & \multicolumn{3}{c}{--} & \multicolumn{3}{c}{--} & \multicolumn{3}{c}{8.33\%} & \multicolumn{3}{c}{--} & \multicolumn{3}{c}{34.52\%} & \multicolumn{3}{c}{5.95\%} \\
Health \& Lifestyle     & \multicolumn{3}{c}{48.21\%} & \multicolumn{3}{c}{--} & \multicolumn{3}{c}{1.03\%} & \multicolumn{3}{c}{1.03\%} & \multicolumn{3}{c}{1.54\%} & \multicolumn{3}{c}{--} & \multicolumn{3}{c}{25.13\%} & \multicolumn{3}{c}{23.08\%}  \\
Medical                 & \multicolumn{3}{c}{60.61\%} & \multicolumn{3}{c}{--} & \multicolumn{3}{c}{--} & \multicolumn{3}{c}{--} & \multicolumn{3}{c}{--} & \multicolumn{3}{c}{--} & \multicolumn{3}{c}{27.27\%} & \multicolumn{3}{c}{12.12\%} \\
News \& Weather         & \multicolumn{3}{c}{40.91\%} & \multicolumn{3}{c}{--} & \multicolumn{3}{c}{4.55\%} & \multicolumn{3}{c}{4.55\%} & \multicolumn{3}{c}{13.64\%} & \multicolumn{3}{c}{--} & \multicolumn{3}{c}{27.27\%} & \multicolumn{3}{c}{9.09\%} \\
Social \& Community     & \multicolumn{3}{c}{44.68\%} & \multicolumn{3}{c}{--} & \multicolumn{3}{c}{--} & \multicolumn{3}{c}{--} & \multicolumn{3}{c}{8.51\%} & \multicolumn{3}{c}{--} & \multicolumn{3}{c}{40.43\%} & \multicolumn{3}{c}{6.38\%} \\
Tools \& Productivity   & \multicolumn{3}{c}{44.00\%} & \multicolumn{3}{c}{3.00\%} & \multicolumn{3}{c}{1.00\%} & \multicolumn{3}{c}{1.00\%} & \multicolumn{3}{c}{8.00\%} & \multicolumn{3}{c}{1.00\%} & \multicolumn{3}{c}{36.00\%} & \multicolumn{3}{c}{6.00\%} \\
Travel \& Navigation    & \multicolumn{3}{c}{68.66\%} & \multicolumn{3}{c}{--} & \multicolumn{3}{c}{--} & \multicolumn{3}{c}{--} & \multicolumn{3}{c}{5.97\%} & \multicolumn{3}{c}{--} & \multicolumn{3}{c}{19.40\%} & \multicolumn{3}{c}{5.97\%} \\
\bottomrule
\end{tabular}
\label{feature_combinations_by_category}
\end{table}

\vspace{2mm}
\begin{center}
     \noindent\fbox{%
    \parbox{0.95\columnwidth}{%
 
      \textbf{(RQ3)} \textbf{How do application popularity (e.g., download number, and user ratings) impact the adoption of BLE security mechanisms in BLE mobile apps?} 
    }%
}
\end{center}
\vspace{2mm}

\begin{figure}[!tb]
	\centering 
	\includegraphics[width=\linewidth]{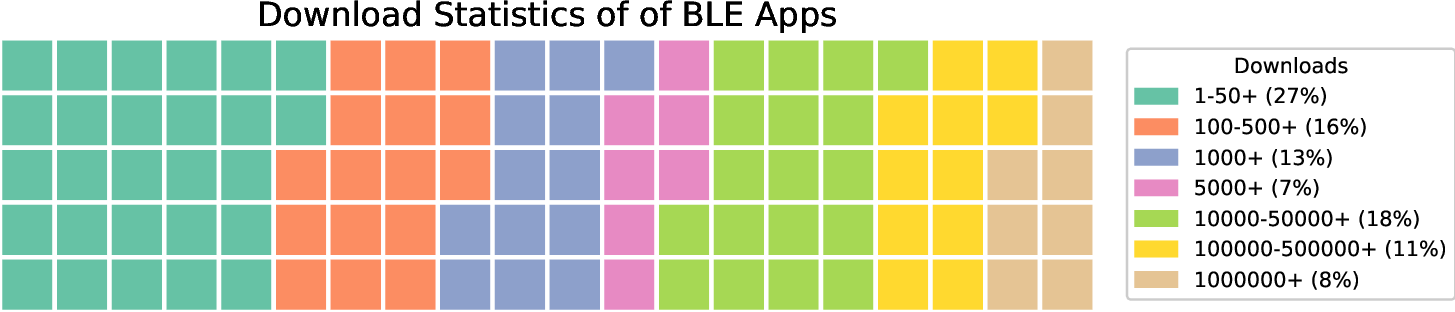}
	\caption{Distribution of Download Numbers for BLE Apps}
	\label{ble_download_stats}
\end{figure}

\begin{figure}[!tb]
	\centering 
	\includegraphics[width=0.45\textwidth]{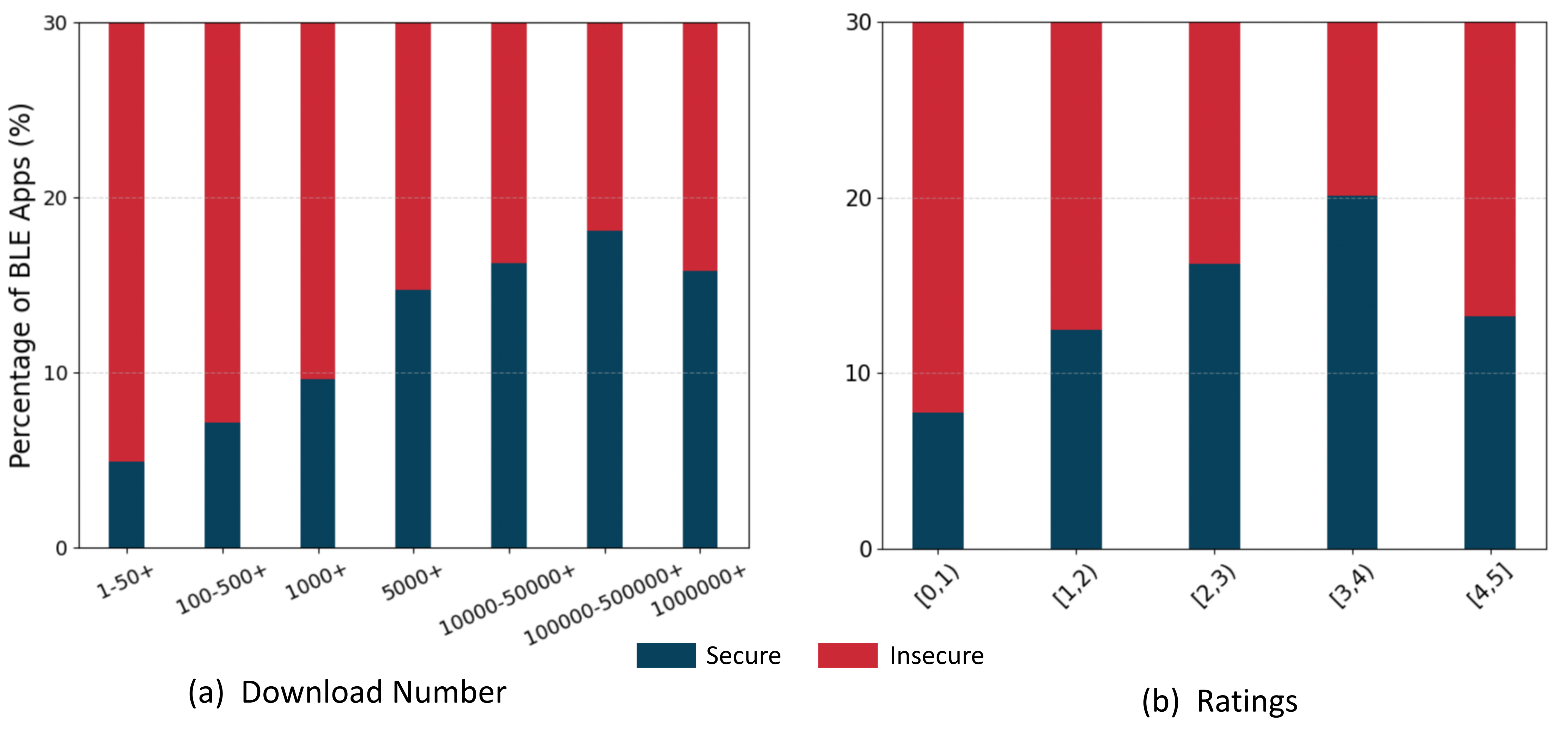}
	\caption{Percentage of Secure/Insecure BLE Apps Across Download Numbers and Ratings}
	\label{ble_security_distribution}
\end{figure}

\noindent 
To examine how BLE security adoption relates to application popularity and user perception, we analyze its distribution across different download ranges and rating intervals.
 \autoref{ble_download_stats} illustrates the distribution of BLE applications across seven download brackets, ranging from $1$–$50+$ to $1,000,000+$. The majority of apps fall into the lower tiers: 27\% in the $1$–$50+$ range, followed by 18\% in $10,000$–$50,000+$ and 16\% in $100$–$500+$. This indicates that most BLE apps serve a relatively small user base, with only 8\% reaching over one million downloads.
\autoref{ble_security_distribution}(a) further explores the relationship between download numbers and security adoption. A clear upward trend emerges: apps with higher download counts are more likely to implement encryption, randomness, and authentication. In contrast, low-download apps frequently omit these essential features. This suggests that widely deployed apps tend to adopt more robust BLE security practices, possibly due to greater development resources or increased scrutiny. Meanwhile,  \autoref{ble_security_distribution}(b) shows how BLE security correlates with user ratings. As app ratings increase, so does the proportion of secure implementations. The [3,4) interval shows the highest security adoption, while low-rated apps (e.g., [0,1) or [1,2)) are predominantly insecure. This pattern suggests that weak security may negatively impact user satisfaction, or that higher-quality apps tend to be more security-conscious.

\vspace{2mm}
\begin{center}
     \noindent\fbox{%
    \parbox{0.95\columnwidth}{%
 
      \textbf{(RQ4)} \textbf{Do developers exhibit consistent security practices across their BLE applications?} 
    }%
}
\end{center}
\vspace{2mm}

\begin{figure*}[!tb]
	\centering 
	\includegraphics[width=0.95\textwidth]{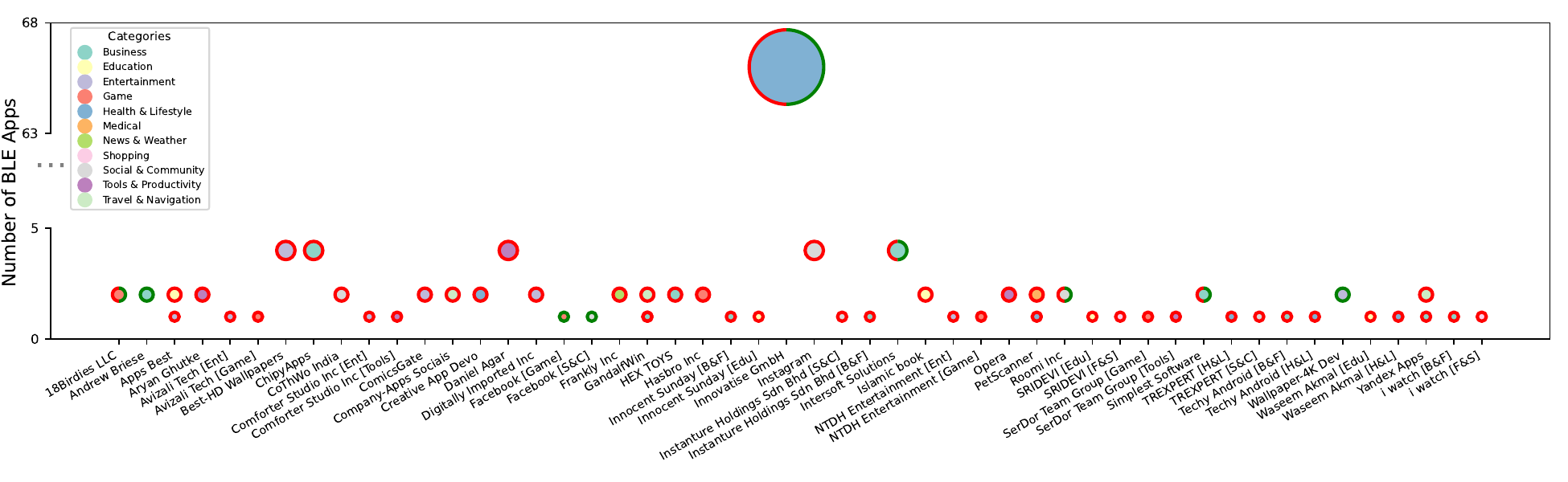}
	\caption{The distribution of apps across developers, categories (by color), and security (by edge color). Bubble size reflects numbers.}
	\label{developer-category-secure}
\end{figure*}

\noindent 
To explore whether BLE security practices are consistent within development entities, we analyze BLE applications at the developer level.
 \autoref{developer-category-secure} presents a bubble chart summarizing BLE app distributions across different developers. In this chart, bubble size indicates the number of BLE apps published by each developer, bubble color represents the app category, and edge color denotes security status: green for secure apps and red for insecure ones. The visualization reveals that most developers have only released a small number of BLE-enabled applications, many of which are insecure. One notable outlier is \textsf{Innovatise GmbH}, which has developed a disproportionately large number of BLE apps, predominantly in the Health \& Lifestyle category. These apps include both secure and insecure implementations, suggesting variability in the developer’s internal security assessment, possibly based on differences in data sensitivity or perceived risk across apps. 
In contrast, a few developers such as \textsf{Facebook} and \textsf{Roomi Inc.} consistently produce secure BLE applications, indicating a likely emphasis on secure development practices or mature compliance workflows. On the other end of the spectrum, developers like \textsf{ComicsGate} and \textsf{Hasbro Inc.} exclusively publish insecure BLE apps, which may reflect a lack of attention to BLE protocol security during development.

\begin{center}
     \noindent\fbox{%
    \parbox{0.95\columnwidth}{%
 
      \textbf{(RQ5)} \textbf{How do BLE security attacks vary across application categories, downloads, and user ratings in mobile apps?} 
    }%
}
\end{center}
\vspace{2mm}

\noindent
To examine the prevalence and patterns of BLE vulnerabilities, we analyze their distribution across app categories, download volumes, and user ratings.
As shown in Table \ref{tab:ble-vuln-distribution}, most BLE apps are insecure, with common vulnerabilities including eavesdropping, spoofing, and MitM.  Only a small fraction of apps implement complete protection. Similarly, \textit{Health \& Lifestyle}, \textit{Business}, and \textit{Medical} apps exhibit relatively higher security coverage. This may be due to their handling of sensitive health data, financial transactions, or personal identifiers, prompting developers to adopt encryption and authentication mechanisms. Meanwhile, security adoption increases with download volume, though overall coverage remains low. Apps with 1–50+ downloads show only 1.25\% secure implementations, while those in the 100,000–500,000+ range reach 5.2\%. However, spoofing and MITM vulnerabilities remain prevalent in all groups, and even high-download apps still suffer from replay risks due to missing freshness mechanisms.
It can also be observed that vulnerabilities persist across all rating intervals. Apps rated [2.0, 3.0) show particularly high replay (23.01\%) and eavesdropping (22.12\%) exposure. Spoofing and MITM attacks remain widespread regardless of rating, highlighting the general absence of robust authentication in BLE implementations.

\begin{table}[htbp]
\centering
\scriptsize
\setlength{\tabcolsep}{2pt}
\caption{Distribution of BLE attacks by Category, Downloads, and Ratings. Cell background intensity reflects the relative severity or prevalence of each security condition, with deeper pink indicating higher percentages. }
\vspace{0.5em}
\begin{tabular}{@{}lccccc@{}}
\toprule
\textbf{Dimension} & \textbf{Secure (\%)} & \multicolumn{4}{c}{\textbf{Insecure (\%)}} \\
\cmidrule(l){3-6}
& & Eavesdropping & Replay & MITM & Spoofing \\
\midrule
\multicolumn{6}{@{}c}{\textit{Category}} \\ \midrule
Education      & \ccell{0.52}{8}{0.52} & \ccell{15.62}{36}{15.62} & \ccell{16.67}{36}{16.67} & \ccell{33.85}{36}{33.85} & \ccell{33.33}{36}{33.33} \\
Business          & \ccell{4.03}{8}{4.03} & \ccell{17.58}{36}{17.58} & \ccell{19.96}{36}{19.96} & \ccell{29.30}{36}{29.30} & \ccell{29.12}{36}{29.12} \\
Tools \& Productivity        & \ccell{1.71}{8}{1.71} & \ccell{16.38}{36}{16.38} & \ccell{19.45}{36}{19.45} & \ccell{31.74}{36}{31.74} & \ccell{30.72}{36}{30.72} \\
Entertainment  & \ccell{1.66}{8}{1.66} & \ccell{14.36}{36}{14.36} & \ccell{17.68}{36}{17.68} & \ccell{33.15}{36}{33.15} & \ccell{33.15}{36}{33.15} \\
Shopping             & \ccell{1.44}{8}{1.44} & \ccell{19.95}{36}{19.95} & \ccell{20.91}{36}{20.91} & \ccell{28.85}{36}{28.85} & \ccell{28.85}{36}{28.85} \\
Game                         & \ccell{1.95}{8}{1.95} & \ccell{16.80}{36}{16.80} & \ccell{19.53}{36}{19.53} & \ccell{30.86}{36}{30.86} & \ccell{30.86}{36}{30.86} \\
Health \& Lifestyle          & \ccell{7.58}{8}{7.58} & \ccell{17.74}{36}{17.74} & \ccell{18.48}{36}{18.48} & \ccell{28.10}{36}{28.10} & \ccell{28.10}{36}{28.10} \\
Medical                      & \ccell{3.92}{8}{3.92} & \ccell{19.61}{36}{19.61} & \ccell{19.61}{36}{19.61} & \ccell{28.43}{36}{28.43} & \ccell{28.43}{36}{28.43} \\
Social \& Community          & \ccell{0.72}{8}{0.72} & \ccell{15.11}{36}{15.11} & \ccell{17.99}{36}{17.99} & \ccell{33.09}{36}{33.09} & \ccell{33.09}{36}{33.09} \\
Travel \& Navigation         & \ccell{1.32}{8}{1.32} & \ccell{20.26}{36}{20.26} & \ccell{22.03}{36}{22.03} & \ccell{28.19}{36}{28.19} & \ccell{28.19}{36}{28.19} \\
News \& Weather              & \ccell{1.59}{8}{1.59} & \ccell{15.87}{36}{15.87} & \ccell{19.05}{36}{19.05} & \ccell{31.75}{36}{31.75} & \ccell{31.75}{36}{31.75} \\
\midrule
\multicolumn{6}{@{}c}{\textit{Download Range}} \\ \midrule
1--50+                       & \ccell{1.25}{8}{1.25} & \ccell{13.30}{36}{13.30} & \ccell{14.68}{36}{14.68} & \ccell{35.46}{36}{35.46} & \ccell{35.32}{36}{35.32} \\
100--500+                    & \ccell{2.14}{8}{2.14} & \ccell{17.31}{36}{17.31} & \ccell{18.59}{36}{18.59} & \ccell{30.98}{36}{30.98} & \ccell{30.98}{36}{30.98} \\
1000+                        & \ccell{2.84}{8}{2.84} & \ccell{18.72}{36}{18.72} & \ccell{20.62}{36}{20.62} & \ccell{29.15}{36}{29.15} & \ccell{28.67}{36}{28.67} \\
5000+                        & \ccell{4.04}{8}{4.04} & \ccell{21.08}{36}{21.08} & \ccell{21.97}{36}{21.97} & \ccell{26.46}{36}{26.46} & \ccell{26.46}{36}{26.46} \\
10000--50000+                & \ccell{4.49}{8}{4.49} & \ccell{19.57}{36}{19.57} & \ccell{21.54}{36}{21.54} & \ccell{27.29}{36}{27.29} & \ccell{27.11}{36}{27.11} \\
100000--500000+              & \ccell{5.20}{8}{5.20} & \ccell{19.57}{36}{19.57} & \ccell{21.71}{36}{21.71} & \ccell{26.91}{36}{26.91} & \ccell{26.61}{36}{26.61} \\
1000000+                     & \ccell{4.20}{8}{4.20} & \ccell{17.65}{36}{17.65} & \ccell{23.53}{36}{23.53} & \ccell{27.31}{36}{27.31} & \ccell{27.31}{36}{27.31} \\
\midrule
\multicolumn{6}{@{}c}{\textit{User Rating}} \\ \midrule
$[0.0, 1.0)$                 & \ccell{2.26}{8}{2.26} & \ccell{17.05}{36}{17.05} & \ccell{18.36}{36}{18.36} & \ccell{31.23}{36}{31.23} & \ccell{31.10}{36}{31.10} \\
$[1.0, 2.0)$                 & \ccell{4.76}{8}{4.76} & \ccell{19.05}{36}{19.05} & \ccell{19.05}{36}{19.05} & \ccell{28.57}{36}{28.57} & \ccell{28.57}{36}{28.57} \\
$[2.0, 3.0)$                 & \ccell{4.42}{8}{4.42} & \ccell{22.12}{36}{22.12} & \ccell{23.01}{36}{23.01} & \ccell{25.66}{36}{25.66} & \ccell{24.78}{36}{24.78} \\
$[3.0, 4.0)$                 & \ccell{5.71}{8}{5.71} & \ccell{17.96}{36}{17.96} & \ccell{21.22}{36}{21.22} & \ccell{27.55}{36}{27.55} & \ccell{27.55}{36}{27.55} \\
$[4.0, 5.0]$                 & \ccell{4.17}{8}{4.17} & \ccell{14.77}{36}{14.77} & \ccell{18.14}{36}{18.14} & \ccell{31.62}{36}{31.62} & \ccell{31.30}{36}{31.30} \\
\bottomrule
\end{tabular}
\label{tab:ble-vuln-distribution}
\end{table}

\begin{center}
     \noindent\fbox{%
    \parbox{0.95\columnwidth}{%
 
      \textbf{(RQ6)} \textbf{How does the security posture of BLE applications evolve over time across different versions?} 
    }%
}
\end{center}
\vspace{2mm}
 
\begin{figure}[!tb]
	\centering 
	\includegraphics[scale=0.42]{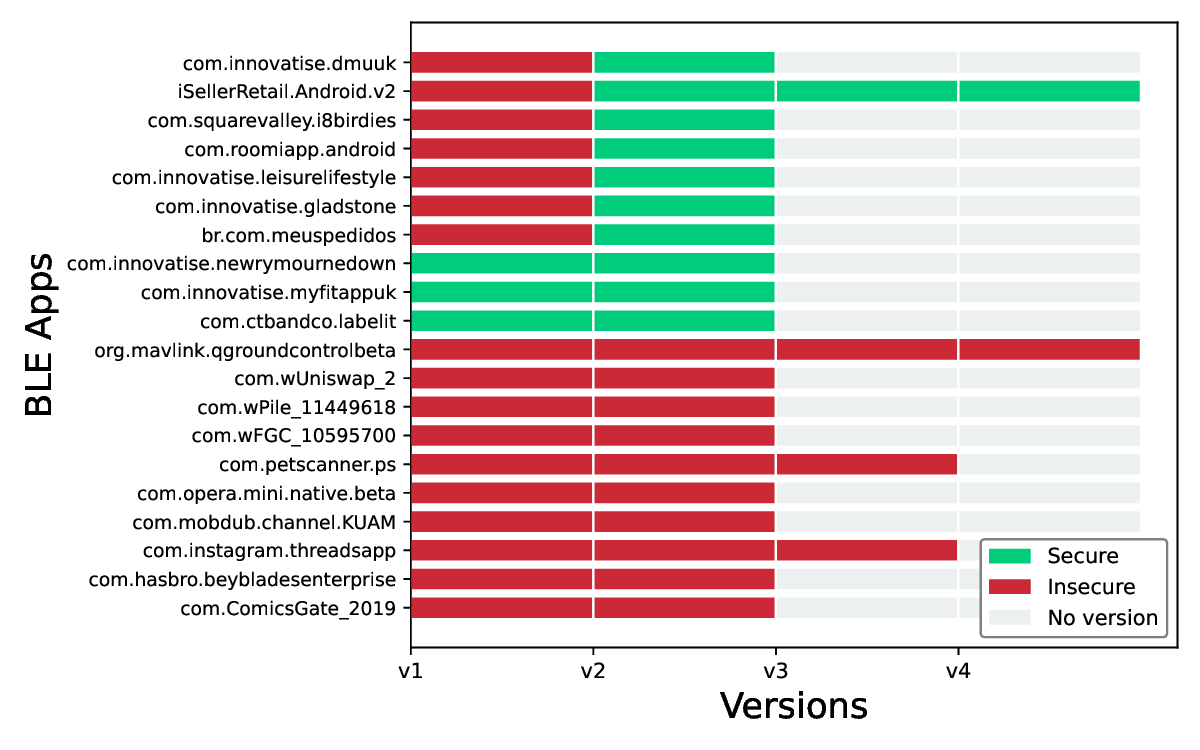}
	\caption{Security Evolution of BLE Applications Across Multiple Versions}
	\label{security_evolution_apk}
\end{figure}

\noindent To examine how BLE security evolves over time, we analyze version-level changes in real-world applications. \autoref{security_evolution_apk} presents the security status of multiple BLE applications across different versions. Each row corresponds to a single app, while the horizontal axis represents its version history. Color coding is used to indicate security status: green for secure versions, red for insecure ones, and gray for versions that are missing or unavailable. Due to data availability constraints, we include only those versions that could be retrieved from our dataset.  The visualization reveals several distinct patterns. Some applications, such as \textit{com.petscanner.ps} and \textit{com.instagram.threadsapp}, remain insecure across all available versions, suggesting a persistent absence of effective security mechanisms throughout their development lifecycle. In our dataset, over 60\% of BLE applications with multiple versions exhibit no observable security improvement over time.  In contrast, a subset of applications (e.g., \textit{com.innovatise.myfitappuk}) show signs of remediation, with earlier insecure versions followed by secure updates. This may reflect developers responding to public vulnerability disclosures, user feedback, or increased regulatory pressure. Meanwhile, a small group of applications maintain consistently secure implementations across all versions, indicating mature and stable development practices. \looseness=-1

\begin{center}
     \noindent\fbox{%
    \parbox{0.95\columnwidth}{%
 
      \textbf{(RQ7)} \textbf{How do missing application-layer protections in BLE devices enable practical attacks?} 
    }%
}
\end{center}
\vspace{2mm}

To evaluate the practical security risks of BLE application-layer flaws, as shown in Table \ref{tab:device_attack_matrix}, we conducted case studies on three popular devices: a smart light, a blood pressure monitor, and a relay switch, each exposing different but equally critical vulnerabilities. For the smart light, we discovered that its BLE commands for adjusting color and brightness were sent without encryption or authentication. By sniffing traffic and analyzing payload differences, we were able to craft and replay arbitrary control commands, taking over device functions without app access.  For the blood pressure monitor, we found that all medical data—systolic/diastolic readings, pulse rates, timestamps, were transmitted in plaintext via BLE notifications. This allows any nearby adversary to silently intercept sensitive health information. Beyond passive eavesdropping, attackers could forge or alter measurement data, misleading patients and even healthcare providers, raising serious privacy and safety concerns in medical environments. The relay switch presented a slightly more complex case, using a three-packet command sequence for device control. However, lacking any cryptographic protection, it was equally vulnerable. By capturing this sequence, we successfully replayed it from a different device to toggle the relay state.

\begin{table}[ht]
\centering
\caption{The Vulnerability Analysis of BLE Devices.
\ding{192}: Eavesdropping, \ding{193}: Replay, \ding{194}: MITM, \ding{195}: Spoofing}
\label{tab:device_attack_matrix}
\scriptsize

\begin{tabular}{llll l}
\toprule
\textbf{Device} & \textbf{Category} & \textbf{Downloads} & \textbf{Release Time} & \textbf{Vulnerabilities} \\
\midrule
Smart Light & Tools   & 10,000,000+ & 2024-08-28 & \ding{192} \ding{193} \ding{194} \ding{195} \\
Relay Switch & Entertainment & 1,000+ & 2021-06-13 & \ding{195} \\
BP Monitor & Medical & 100+ & 2023-06-07 & \ding{192} \ding{193} \ding{194} \ding{195} \\
\bottomrule
\end{tabular}
\end{table}

\section{Discussion}\label{sec7}

\noindent\textbf{Lessons for Software Engineering.} 
A key insight from our work is that LLMs are most effective not when used to directly detect vulnerabilities, but when positioned as semantic translators that bridge the gap between code and formal representations.  This design principle offers a generalizable strategy for many software engineering tasks. For example, in automated testing, LLMs could translate code snippets or docstrings into symbolic path constraints or test specifications, which can then be executed or analyzed using fuzzers or symbolic execution engines. In specification mining, natural-language comments or implementation code can be converted into formal contracts (e.g., pre/postconditions) to support runtime verification or static analysis. For log analysis, LLMs may help translate raw system logs into finite-state behavioral models or temporal properties, enabling the use of model checking or anomaly detection tools. Similarly, for program understanding, LLMs could extract abstract representations such as control-flow graphs or dataflow summaries, which are then fed into performance profilers, security analyzers, or refactoring engines.

\vspace{1mm}
\noindent\textbf{Ethical Considerations.}
All experiments in this work were conducted in accordance with ethical guidelines for responsible security research. We tested only BLE devices that we owned, and all experiments were performed in isolated environments without affecting third-party users or systems. When real-world vulnerabilities were identified in commercial products, we avoided disclosing exploit details and followed responsible disclosure practices when applicable.   Furthermore, our analysis of commercial applications was limited to static inspection of publicly available APKs and passive BLE traffic capture, without violating terms of service or tampering with proprietary systems. 

\section{Related Work}\label{sec8}

We review prior work on BLE security, LLM-based analysis, and LLM-assisted formal verification.  Unlike prior work that uses LLMs directly for vulnerability detection or synthesis, our approach leverages LLMs as semantic translators to generate formal models, enabling precise and auditable security verification through established symbolic tools.

\vspace{1mm}
\noindent\textbf{BLE Security.}
BLE protocols are known to suffer from vulnerabilities in pairing, encryption, and reconnection procedures \cite{garbelini2020sweyntooth,wu2020blesa,wu2024finding,jangid2023extrapolating,che2024blueswat,9916280}. Attacks such as SweynTooth and BLESA demonstrate that flawed implementations can lead to device crashes or unauthorized reconnections \cite{garbelini2020sweyntooth,wu2020blesa}. To address this, formal tools like \textsf{ProVerif} have been applied to verify authentication logic \cite{wu2024finding,jangid2023extrapolating}, and lightweight protocols have been proposed to improve security under resource constraints \cite{che2024blueswat}. 
Pairing remains a major weak point. Just Works and static PIN schemes allow MITM or brute-force attacks \cite{zhang2020breaking}, while MiniBLE revealed hard-coded PINs in many WeChat BLE apps \cite{zhang2023minible}. Defenses include Secure Connections mode and ephemeral key use.

\vspace{1mm}

\noindent\textbf{LLMs for Security and Formal Verification}
LLMs have been explored for vulnerability detection, repair, and adversarial defense \cite{du2024vul,10462177,thapa2022transformer,10301302,li2024drattack,guo2024cold,li2023deepinception,chen2023can,li2024attention,pearce2023examining,yan2025protecting,yao2024survey}. They show promise over traditional methods in certain tasks \cite{du2024vul}, but face challenges like high false positives and limited semantic precision \cite{10301302}. Adversarial prompts and jailbreaks also pose risks \cite{li2024drattack,guo2024cold,li2023deepinception}, prompting research on filtering and prompt hardening \cite{chen2023can}. 
Recent work combines LLMs with formal tools for tasks like assertion generation, protocol modeling, and program synthesis \cite{hassan2024llm,ferrag2025securefalcon,sevenhuijsen2024effectiveness,liu2024domain,kumar2024generative,10458667,zhou2024don-verify}. In this setting, LLMs generate candidate artifacts, while SMT solvers or model checkers verify correctness, reducing hallucination risk. Some methods embed formal constraints into LLM prompting to improve generation quality \cite{lee2025veriplan,li2024formal,lin2024fvel}, and others use LLMs to infer security goals for cryptographic protocols \cite{curaba2cryptoformaleval}.

\section{Conclusion}\label{sec9}

This paper introduces \tool, a novel system that uses LLMs as semantic translators to bridge the gap between real-world BLE application logic and formal verification tools. By shifting the focus from direct vulnerability detection to model generation, our approach overcomes the practical barriers of formal analysis, making it scalable, interpretable, and applicable to real-world apps. Through a large-scale study of 1,050 BLE Android applications, we reveal that the majority lack fundamental protections such as encryption, freshness, and authentication, exposing users to widespread risks like spoofing and replay attacks. Our findings also uncover strong correlations between BLE security and app popularity, domain, and developer consistency.

\bibliographystyle{IEEEtran}
\bibliography{TIFS-BLE}

\begin{thebibliography}{10}
\providecommand{\url}[1]{#1}
\csname url@samestyle\endcsname
\providecommand{\newblock}{\relax}
\providecommand{\bibinfo}[2]{#2}
\providecommand{\BIBentrySTDinterwordspacing}{\spaceskip=0pt\relax}
\providecommand{\BIBentryALTinterwordstretchfactor}{4}
\providecommand{\BIBentryALTinterwordspacing}{\spaceskip=\fontdimen2\font plus
\BIBentryALTinterwordstretchfactor\fontdimen3\font minus
  \fontdimen4\font\relax}
\providecommand{\BIBforeignlanguage}[2]{{%
\expandafter\ifx\csname l@#1\endcsname\relax
\typeout{** WARNING: IEEEtran.bst: No hyphenation pattern has been}%
\typeout{** loaded for the language `#1'. Using the pattern for}%
\typeout{** the default language instead.}%
\else
\language=\csname l@#1\endcsname
\fi
#2}}
\providecommand{\BIBdecl}{\relax}
\BIBdecl

\bibitem{zuo2019automatic}
C.~Zuo, H.~Wen, Z.~Lin, and Y.~Zhang, ``Automatic fingerprinting of vulnerable
  {BLE} {IoT} devices with static {UUIDs} from mobile apps,'' in
  \emph{Proceedings of the 2019 ACM SIGSAC Conference on Computer and
  Communications Security}, 2019, pp. 1469--1483.

\bibitem{zhang2020bless}
Y.~Zhang, J.~Weng, Z.~Ling, B.~Pearson, and X.~Fu, ``{BLESS}: A {BLE}
  application security scanning framework,'' in \emph{IEEE INFOCOM 2020-IEEE
  Conference on Computer Communications}.\hskip 1em plus 0.5em minus
  0.4em\relax IEEE, 2020, pp. 636--645.

\bibitem{wu2020blesa}
J.~Wu, Y.~Nan, V.~Kumar, D.~J. Tian, A.~Bianchi, M.~Payer, and D.~Xu,
  ``{BLESA}: Spoofing attacks against reconnections in bluetooth low energy,''
  in \emph{14th USENIX Workshop on Offensive Technologies (WOOT 20)}, 2020.

\bibitem{zhang2020breaking}
Y.~Zhang, J.~Weng, R.~Dey, Y.~Jin, Z.~Lin, and X.~Fu, ``Breaking secure pairing
  of bluetooth low energy using downgrade attacks,'' in \emph{29th USENIX
  Security Symposium (USENIX Security 20)}, 2020, pp. 37--54.

\bibitem{wen2020firmxray}
H.~Wen, Z.~Lin, and Y.~Zhang, ``{FirmXRay}: Detecting bluetooth link layer
  vulnerabilities from bare-metal firmware,'' in \emph{Proceedings of the 2020
  ACM SIGSAC conference on computer and communications security}, 2020, pp.
  167--180.

\bibitem{woolley2019bluetooth}
M.~Woolley, ``Bluetooth core specification v5. 1,'' \emph{Bluetooth Special
  Interest Group}, 2019.

\bibitem{vaswani2017attention}
A.~Vaswani, N.~Shazeer, N.~Parmar, J.~Uszkoreit, L.~Jones, A.~N. Gomez,
  {\L}.~Kaiser, and I.~Polosukhin, ``Attention is all you need,''
  \emph{Advances in neural information processing systems}, vol.~30, 2017.

\bibitem{li2024attention}
Y.~Li, X.~Li, H.~Wu, Y.~Zhang, X.~Cheng, S.~Zhong, and F.~Xu, ``Attention is
  all you need for {LLM-based} code vulnerability localization,'' \emph{arXiv
  preprint arXiv:2410.15288}, 2024.

\bibitem{yao2024survey}
Y.~Yao, J.~Duan, K.~Xu, and et~al., ``A survey on large language model ({LLM})
  security and privacy: The good, the bad, and the ugly,''
  \emph{High-Confidence Computing}, p. 100211, 2024.

\bibitem{10606356}
S.~Fakhoury, A.~Naik, G.~Sakkas, S.~Chakraborty, and S.~K. Lahiri, ``Llm-based
  test-driven interactive code generation: User study and empirical
  evaluation,'' \emph{IEEE Transactions on Software Engineering}, vol.~50,
  no.~9, pp. 2254--2268, 2024.

\bibitem{930138}
B.~Blanchet, ``An efficient cryptographic protocol verifier based on prolog
  rules,'' in \emph{Proceedings. 14th IEEE Computer Security Foundations
  Workshop, 2001.}, 2001, pp. 82--96.

\bibitem{yin2024multitask}
X.~Yin, C.~Ni, and S.~Wang, ``Multitask-based evaluation of open-source {LLM}
  on software vulnerability,'' \emph{IEEE Transactions on Software
  Engineering}, vol.~50, no.~11, pp. 3071--3087, 2024.

\bibitem{steenhoek2024err}
B.~Steenhoek, M.~M. Rahman, M.~K. Roy, M.~S. Alam, H.~Tong, S.~Das, E.~T. Barr,
  and W.~Le, ``To {Err} is machine: Vulnerability detection challenges {LLM}
  reasoning,'' \emph{arXiv preprint arXiv:2403.17218}, 2024.

\bibitem{ding2024vulnerability}
Y.~Ding, Y.~Fu, O.~Ibrahim, C.~Sitawarin, X.~Chen, B.~Alomair, D.~Wagner,
  B.~Ray, and Y.~Chen, ``Vulnerability detection with code language models: How
  far are we?'' in \emph{2025 IEEE/ACM 47th International Conference on
  Software Engineering (ICSE)}.\hskip 1em plus 0.5em minus 0.4em\relax IEEE
  Computer Society, 2024, pp. 469--481.

\bibitem{allix2016androzoo}
K.~Allix, T.~F. Bissyand{\'e}, J.~Klein, and Y.~Le~Traon, ``Androzoo:
  Collecting millions of android apps for the research community,'' in
  \emph{Proceedings of the 13th international conference on mining software
  repositories}, 2016, pp. 468--471.

\bibitem{sivakumaran2019study}
P.~Sivakumaran and J.~Blasco, ``A study of the feasibility of co-located app
  attacks against {BLE} and a {Large-Scale} analysis of the current
  {Application-Layer} security landscape,'' in \emph{28th USENIX Security
  Symposium (USENIX Security 19)}, 2019, pp. 1--18.

\bibitem{garbelini2020sweyntooth}
M.~E. Garbelini, C.~Wang, S.~Chattopadhyay, S.~Sumei, and E.~Kurniawan,
  ``{SweynTooth}: unleashing mayhem over bluetooth low energy,'' in \emph{2020
  USENIX Annual Technical Conference (USENIX ATC 20)}, 2020, pp. 911--925.

\bibitem{wu2024finding}
J.~Wu, P.~Traynor, D.~Xu, D.~J. Tian, and A.~Bianchi, ``Finding traceability
  attacks in the bluetooth low energy specification and its implementations,''
  in \emph{33rd USENIX Security Symposium (USENIX Security 24)}, 2024, pp.
  4499--4516.

\bibitem{jangid2023extrapolating}
M.~K. Jangid, Y.~Zhang, and Z.~Lin, ``Extrapolating formal analysis to uncover
  attacks in bluetooth passkey entry pairing.'' in \emph{NDSS}, 2023.

\bibitem{che2024blueswat}
X.~Che, Y.~He, X.~Feng, K.~Sun, K.~Xu, and Q.~Li, ``{BlueSWAT}: A lightweight
  state-aware security framework for bluetooth low energy,'' in
  \emph{Proceedings of the 2024 on ACM SIGSAC Conference on Computer and
  Communications Security}, 2024, pp. 2087--2101.

\bibitem{9916280}
L.~David, A.~Hassidim, Y.~Matias, M.~Yung, and A.~Ziv, ``Eddystone-eid: Secure
  and private infrastructural protocol for ble beacons,'' \emph{IEEE
  Transactions on Information Forensics and Security}, vol.~17, pp. 3877--3889,
  2022.

\bibitem{zhang2023minible}
Z.~Zhang, J.~Du, W.~Diao, and J.~Wu, ``{MiniBLE}: Exploring insecure {BLE}
  {API} usages in mini-programs,'' in \emph{Proceedings of the ACM Workshop on
  Secure and Trustworthy Superapps}, 2023, pp. 18--22.

\bibitem{du2024vul}
X.~Du, G.~Zheng, K.~Wang, J.~Feng, W.~Deng, M.~Liu, B.~Chen, X.~Peng, T.~Ma,
  and Y.~Lou, ``{Vul-RAG}: Enhancing {LLM}-based vulnerability detection via
  knowledge-level {RAG},'' \emph{CoRR}, 2024.

\bibitem{10462177}
B.~Ahmad, S.~Thakur, B.~Tan, R.~Karri, and H.~Pearce, ``On hardware security
  bug code fixes by prompting large language models,'' \emph{IEEE Transactions
  on Information Forensics and Security}, vol.~19, pp. 4043--4057, 2024.

\bibitem{thapa2022transformer}
C.~Thapa, S.~I. Jang, M.~E. Ahmed, S.~Camtepe, J.~Pieprzyk, and S.~Nepal,
  ``Transformer-based language models for software vulnerability detection,''
  in \emph{Proceedings of the 38th Annual Computer Security Applications
  Conference}, 2022, pp. 481--496.

\bibitem{10301302}
M.~D. Purba, A.~Ghosh, B.~J. Radford, and B.~Chu, ``Software vulnerability
  detection using large language models,'' in \emph{2023 IEEE 34th
  International Symposium on Software Reliability Engineering Workshops
  (ISSREW)}, 2023, pp. 112--119.

\bibitem{li2024drattack}
X.~Li, R.~Wang, M.~Cheng, T.~Zhou, and C.-J. Hsieh, ``{DrAttack}: Prompt
  decomposition and reconstruction makes powerful {LLMs} jailbreakers,'' in
  \emph{Findings of the Association for Computational Linguistics: EMNLP 2024},
  2024, pp. 13\,891--13\,913.

\bibitem{guo2024cold}
X.~Guo, F.~Yu, H.~Zhang, L.~Qin, and B.~Hu, ``{COLD-attack}: Jailbreaking
  {LLMs} with stealthiness and controllability,'' in \emph{Proceedings of the
  41st International Conference on Machine Learning}, 2024, pp.
  16\,974--17\,002.

\bibitem{li2023deepinception}
X.~Li, Z.~Zhou, J.~Zhu, J.~Yao, T.~Liu, and B.~Han, ``{DeepInception}:
  Hypnotize large language model to be jailbreaker,'' in \emph{Neurips Safe
  Generative AI Workshop 2024}, 2024.

\bibitem{chen2023can}
Y.~Chen, A.~Arunasalam, and Z.~B. Celik, ``Can large language models provide
  security \& privacy advice? measuring the ability of llms to refute
  misconceptions,'' in \emph{Proceedings of the 39th Annual Computer Security
  Applications Conference}, 2023, pp. 366--378.

\bibitem{pearce2023examining}
H.~Pearce, B.~Tan, B.~Ahmad, R.~Karri, and B.~Dolan-Gavitt, ``Examining
  zero-shot vulnerability repair with large language models,'' in \emph{2023
  IEEE Symposium on Security and Privacy (SP)}.\hskip 1em plus 0.5em minus
  0.4em\relax IEEE, 2023, pp. 2339--2356.

\bibitem{yan2025protecting}
B.~Yan, K.~Li, M.~Xu, and et~al., ``On protecting the data privacy of large
  language models ({LLMs}) and {LLM} agents: A literature review,''
  \emph{High-Confidence Computing}, p. 100300, 2025.

\bibitem{hassan2024llm}
M.~Hassan, S.~Ahmadi-Pour, K.~Qayyum, C.~K. Jha, and R.~Drechsler,
  ``{LLM-guided} formal verification coupled with mutation testing,'' in
  \emph{2024 Design, Automation \& Test in Europe Conference \& Exhibition
  (DATE)}.\hskip 1em plus 0.5em minus 0.4em\relax IEEE, 2024, pp. 1--2.

\bibitem{ferrag2025securefalcon}
M.~A. Ferrag, A.~Battah, N.~Tihanyi, R.~Jain, D.~Maimu{\c{t}}, F.~Alwahedi,
  T.~Lestable, N.~S. Thandi, A.~Mechri, M.~Debbah \emph{et~al.},
  ``{SecureFalcon}: Are we there yet in automated software vulnerability
  detection with {LLMs}?'' \emph{IEEE Transactions on Software Engineering},
  2025.

\bibitem{sevenhuijsen2024effectiveness}
M.~Sevenhuijsen, ``Effectiveness of large language models to generate formally
  verified {C} code,'' 2024.

\bibitem{liu2024domain}
M.~Liu, M.~Kang, G.~B. Hamad, S.~Suhaib, and H.~Ren, ``Domain-adapted {LLMs}
  for {VLSI} design and verification: A case study on formal verification,'' in
  \emph{2024 IEEE 42nd VLSI Test Symposium (VTS)}.\hskip 1em plus 0.5em minus
  0.4em\relax IEEE, 2024, pp. 1--4.

\bibitem{kumar2024generative}
A.~Kumar and D.~N. Gadde, ``Generative {AI} augmented induction-based formal
  verification,'' in \emph{2024 IEEE 37th International System-on-Chip
  Conference (SOCC)}.\hskip 1em plus 0.5em minus 0.4em\relax IEEE, 2024, pp.
  1--2.

\bibitem{10458667}
R.~Kande, H.~Pearce, B.~Tan, B.~Dolan-Gavitt, S.~Thakur, R.~Karri, and
  J.~Rajendran, ``({Security}) assertions by large language models,''
  \emph{IEEE Transactions on Information Forensics and Security}, vol.~19, pp.
  4374--4389, 2024.

\bibitem{zhou2024don-verify}
J.~P. Zhou, C.~Staats, W.~Li, C.~Szegedy, K.~Q. Weinberger, and Y.~Wu, ``Don't
  trust: Verify-grounding {LLM} quantitative reasoning with
  autoformalization,'' in \emph{ICLR}, 2024.

\bibitem{lee2025veriplan}
C.~P. Lee, D.~Porfirio, X.~J. Wang, K.~C. Zhao, and B.~Mutlu, ``{VeriPlan}:
  Integrating formal verification and {LLMs} into end-user planning,'' in
  \emph{Proceedings of the 2025 CHI Conference on Human Factors in Computing
  Systems}, 2025, pp. 1--19.

\bibitem{li2024formal}
Z.~Li, W.~Hua, H.~Wang, H.~Zhu, and Y.~Zhang, ``{Formal-LLM}: Integrating
  formal language and natural language for controllable {LLM-based} agents,''
  \emph{CoRR}, 2024.

\bibitem{lin2024fvel}
X.~Lin, Q.~Cao, Y.~Huang, H.~Wang, J.~Lu, Z.~Liu, L.~Song, and X.~Liang,
  ``{FVEL}: Interactive formal verification environment with large language
  models via theorem proving,'' \emph{Advances in Neural Information Processing
  Systems}, vol.~37, pp. 54\,932--54\,946, 2024.

\bibitem{curaba2cryptoformaleval}
C.~Curaba, D.~Denis, and A.~Minisini, ``{CryptoFormalEval}: Integrating large
  language models and formal verification for automated cryptographic protocol
  vulnerability detection,'' in \emph{The First Workshop on System-2 Reasoning
  at Scale, NeurIPS'24}, 2024.

\end{thebibliography}

\vfill

\end{document}